\documentclass[11pt,a4paper]{article}
\pdfoutput=1
\usepackage{jheppub}

\usepackage{amsmath,hyperref,amssymb}
\usepackage{graphicx}
\usepackage{dcolumn}
\usepackage{bm}
\usepackage{verbatim}
\usepackage{tabularx}
\usepackage{slashed}
\usepackage{cancel}
\usepackage[boxsize=2em,aligntableaux=center]{ytableau}
\usepackage{float}
\usepackage{color}

\def\be{\begin{equation}}
\def\ee{\end{equation}}
\def\bea{\begin{eqnarray}}
\def\eea{\end{eqnarray}}

\newcommand{\cm}{{\rm cm}}
\newcommand{\km}{{\rm km}}

\newcommand{\Mev}{{\rm MeV}}

\newcommand{\mev}{{\rm MeV}}

\newcommand{\gprime}{{\gamma^\prime}}

\title{Observable signatures of dark photons from supernovae}

\author[a]{William DeRocco,}
\author[a]{Peter W. Graham,}
\author[b,c]{Daniel Kasen,}
\author[a,d]{Gustavo Marques-Tavares,}
\author[e]{Surjeet Rajendran,}

\affiliation[a]{Stanford Institute for Theoretical Physics, \\
Stanford University, Stanford, CA 94305, USA}
\affiliation[b]{Berkeley Center for Theoretical Physics, Department of Physics, \\
University of California, Berkeley, CA 94720, USA}
\affiliation[c]{Lawrence Berkeley National Laboratory, Berkeley, CA 94720, USA}
\affiliation[d]{Maryland Center for Fundamental Physics, Department of Physics, \\
University of Maryland, College Park, MD 20742}
\affiliation[e]{Berkeley Center for Theoretical Physics, Department of Physics, \\
University of California, Berkeley, CA 94720, USA}

\emailAdd{wderocco@stanford.edu}
\emailAdd{pwgraham@stanford.edu}
\emailAdd{kasen@berkeley.edu}
\emailAdd{gusmt@umd.edu}
\emailAdd{surjeet@berkeley.edu}

\vspace*{1cm}

\abstract{A dark photon is a well-motivated new particle which, as a component of an associated dark sector, could explain dark matter.  One strong limit on dark photons arises from excessive cooling of supernovae.  We point out that even at couplings where too few dark photons are produced in supernovae to violate the cooling bound, they can be observed directly through their decays.
Supernovae produce dark photons which decay to positrons, giving a signal in the 511 keV annihilation line observed by SPI/INTEGRAL.  Further, prompt gamma-ray emission by these decaying dark photons gives a signal for gamma-ray telescopes.  Existing GRS observations of SN1987a already constrain this, and a future nearby SN could provide a detection. Finally, dark photon decays from extragalactic SN would produce a diffuse flux of gamma rays observable by detectors such as SMM and HEAO-1.  Together these observations can probe dark photon couplings several orders of magnitude beyond current constraints for masses of roughly 1 - 100 MeV.}

\arxivnumber{1901.08596}

\begin{document}
\maketitle

\section{Introduction}

New degrees of freedom beyond the Standard Model could be hiding at energies significantly below the weak scale if the new particles have very weak couplings to the Standard Model (SM). Such scenarios are often known as ``dark sectors'' and naturally appear in many extensions of the SM. In particular, dark matter, which represents more than $80\%$ of the matter density of the universe and whose identity remain one of the biggest mysteries in physics, could be part of a dark sector.

The new particles in dark sector models can only interact with the SM via a mediator. Though there is a rich experimental program searching for dark sectors and the corresponding mediators~(see e.g. Refs.~\cite{Essig:2013lka,Alexander:2016aln}), if the mediators only interact very weakly with SM degrees of freedom, they could evade these searches and remain hidden. Astrophysical probes can greatly enhance the sensitivity reach for such scenarios, trading the precision associated with the controlled environment of a lab for the enormous densities and temperatures of stars~\cite{Raffelt:1996wa}.

One of the simplest and most well-motivated mediators is the dark photon, a $U(1)'$ gauge boson kinetically mixed with the SM photon~\cite{Holdom:1985ag}. In this paper, we focus on the scenario in which the dark photon decays purely into electron-positron pairs, which is the natural expectation if the dark photon is lighter than the other particles in the dark sector and is above the electron-positron mass threshold.  At the energies we are interested in, the relevant terms in the Lagrangian containing the dark photon are:
\be
\mathcal{L} \supset \frac{1}{2} m' A'_{\mu} A'^{\mu} - \frac{1}{4} F'_{\mu\nu}F'^{\mu\nu} - \frac{\epsilon}{2}F'_{\mu\nu}F^{\mu\nu},
\ee
where $A'$ is the dark photon and $\epsilon$ is the mixing parameter.

A strong constraint in this model parameter space comes from supernovae cooling considerations using the ``Raffelt criterion,'' which simply states that if any new particle could transport energy out of the SN core more efficiently than the neutrinos, it would significantly alter the cooling timescale of the proto-neutron star in conflict with existing measurements of SN 1987a~\cite{Raffelt:1996wa}. Quantitatively, this amounts to asserting that if a new particle carries more than $3\times10^{52}$ erg/s out of the neutrinosphere, it is excluded. This bound was recently updated in Refs.~\cite{Chang:2016ntp,Hardy:2016kme} to include finite temperature and density effects in the dark photon production which lead to significant changes compared to the original calculations. 

Dark photons produced within the core of a supernova can escape the progenitor star before decaying into electrons and positrons. In this work we point out that even in the regime where too few dark photons escape to yield a cooling bound, observable signatures are still produced by their decays. We show that prompt gamma-ray emission from the next galactic supernova would allow for the discovery of the dark photon. Furthermore, we use current measurements to constrain the mixing parameter in the regime where the coupling is too weak to lead to a cooling bound. We apply three different constraints to cover all of our parameter space. These are: (1) requiring that the decays of dark photons produced in supernovae do not lead to an overabundance of galactic positrons, (2) requiring that prompt gamma-ray emission from dark photon decays is not in conflict with observations of SN1987a, and (3) requiring that any plasmas formed by the decay products do not lead to an excess of extragalactic gamma rays over current measurements.

\section{Dark Photon Production}

In this section, we reproduce the standard computation for dark photon emission from a supernova. For a discussion of new bounds, see Sections III through V.

The temperatures inside the SN core reach $\sim 30$ MeV, hence dark photons of masses less than about 100 MeV can be thermally produced in significant numbers within the SN. Plasma effects also cause resonant production of dark photons at sufficiently high electron densities.
We follow the calculation described in Ref.~\cite{Chang:2016ntp} to determine the total flux.

The differential number flux of dark photons from the SN is given by 
\be
\label{Eq: dndv}
\frac{dN}{dV~dt} = \int d\omega \frac{dN}{dV~dt~d\omega}= \int \frac{d\omega~\omega^2 v}{2 \pi^2} e^{-\omega/T} \left(\Gamma'_{\text{abs},L} +2\Gamma'_{\text{abs},T}\right)
\ee
where $T$ is the temperature, $v$ is the velocity, and $\Gamma'_{\text{abs},L/T} $ is the absorptive width of the dark photon for the longitudinal/transverse modes. We consider only inverse bremsstrahlung, which is the dominant absorptive process inside the core. The absorptive widths are then given by
\be
\Gamma'_{\text{ibr}, L|T} = \frac{32}{3\pi}\frac{\alpha(\epsilon_m)^2_{L|T} n_n n_p}{\omega^3}\left(\frac{\pi T}{m_N}\right)^{3/2}\langle\sigma_{np}^{(2)}(T)\rangle \left(\frac{m'^2}{\omega^2}\right)_L
\ee
where $n_n$ and $n_p$ are the neutron and proton number densities respectively, $m_N = 938$ MeV, $\langle\sigma_{np}^{(2)}(T)\rangle$ is the averaged neutron-proton dipole scattering cross-section (taken from~\cite{Rrapaj:2015wgs}), $(\epsilon_m)^2_{L|T}$ is the in-medium mixing angle, and the final term is denoted with a subscript $L$ to indicate that it is only included for the longitudinal mode.

Plasma effects cause the mixing parameter $\epsilon$ to change in medium. We can find this $(\epsilon_m)^2_{L|T}$ by using
\be
(\epsilon_m)^2_{L|T} = \frac{\epsilon^2}{(1 - \text{Re}~\Pi_{L|T}/m'^2)^2 +(\text{Im}~\Pi/m'^2)^2)}
\ee
with $\Pi$ the photon polarization tensor~\cite{An:2013yfc}. The real part of the polarization tensor for the two modes is given by
\begin{align}
&\text{Re}~\Pi_L = \frac{3\omega_p^2}{v^2}(1-v^2)\left[\frac{1}{2v}\ln\left(\frac{1+v}{1-v}\right)-1\right] &\\
&\text{Re}~\Pi_T = \frac{3\omega_p^2}{2v^2}\left[1-\frac{1-v^2}{2v}\ln\left(\frac{1+v}{1-v}\right)\right]&
\end{align}
with $v = k/\omega$ and $\omega_p$ the plasma frequency, which for a gas of degenerate electrons is given by
\be
\omega_p^2 = \frac{4\pi\alpha_{\text{EM}}n_e}{\sqrt{m_e^2 + (3\pi^2 n_e)^{2/3}}}
\ee
where $n_e$ denotes the number density of electrons~\cite{Braaten:1993jw}.

Within the SN, the SM photons are in thermal equilibrium. Hence the imaginary part of the polarization tensor just becomes
\be
\text{Im}~\Pi_{L, T} = -\omega \left(1 - e^{-\omega/T}\right)\Gamma_{\text{abs}|L, T}
\ee
where $\Gamma_{\text{abs}|L, T}$ is the absorptive width of the \textit{Standard Model photon}, taken to be
\be
\Gamma_{\text{ibr}, L|T} = \frac{32\alpha}{3\pi}\frac{ n_n n_p}{\omega^3}\left(\frac{\pi T}{m_N}\right)^{3/2}\langle\sigma_{np}^{(2)}(T)\rangle \left(\frac{m'^2}{\omega^2}\right)_L.
\ee
In other words, $\Gamma'_{\text{abs}} = (\epsilon_m)^2_{L|T}\Gamma_{\text{abs}}$.

By using all of the above relations, we can now determine $dN/dV$ as a function of radius. The radial dependence comes in the form of the radially-dependent parameters $n_n(r)$, $n_p(r)$, $n_e(r)$, and $T(r)$. We use the same fiducial profile as Ref.~\cite{Chang:2016ntp}, which has the functional form
\be
\label{eq:density}
\rho(r) = \rho_c \times
\begin{cases}
1 + k_{\rho}(1-r/R_c) & r < R_c \\
(r/R_c)^{-\nu} & r \geq R_c
\end{cases}
\ee
\be
T(r) = T_c \times
\begin{cases}
1 + k_{T}(1-r/R_c) & r < R_c \\
(r/R_c)^{-\nu/3} & r \geq R_c
\end{cases}
\ee
 with $k_{\rho} = 0.2$, $k_T = -0.5$, $\nu = 5$, $R_c = 10~\km$, $\rho_c = 3\times10^{14}~{\rm g/cm^3}$, $T_c = 30~\Mev$, and constant $Y = 0.3$. We have chosen to use this profile since we find that it produces slightly more conservative bounds in comparison to other profiles, though we find that the bounds change little even under a large variation in the profile. (See Appendix A.)

We perform the integral $\int \frac{dN}{dt~dV} ~dV$ over these radial profiles to compute the total dark photon flux. This flux is dominated by regions within or near the core, so the upper limit of this integral can be taken to be $\sim50$ km without affecting the total flux. In the region of parameter space we are focusing on, the mean free path of the dark photon is sufficiently large such that we can safely neglect possible reabsorption near the core.\footnote{It should be noted that the density profile in Eq.~\ref{eq:density} is only applicable near the core, as one would expect an exponential decrease in density at large radii. However, since in the regions of parameter space of interest, the dark photons are produced in the core and escape without interacting with the SN overburden, the dependence at large radii does not affect the analysis.} 

\section{Positron Bounds}

In this section, we show that positrons can provide a direct signature of dark photons emitted by supernovae and discuss their behavior in a variety of regimes.

\subsection{Galactic positrons and supernova progenitors}

The results from the SPI gamma ray spectrometer on the INTEGRAL satellite have greatly improved the measurements of the galactic 511 keV gamma ray flux~\cite{Siegert:2015knp}, which in turn have allowed for more accurate estimates of the galactic positron annihilation rate. Estimates of this rate vary depending on the particular galactic model used, but suggest a rate no larger than $\sim4\times10^{43}~\text{s}^{-1}$~\cite{Prantzos:2010wi}. Assuming that positron production and annihilation are in equilibrium, this can be taken as the galactic positron production rate as well. If some set of astrophysical sources were to produce positrons in excess of this rate, the resulting positrons would produce an excess of 511 keV photons in the galaxy in disagreement with INTEGRAL's result. Using only the bound from the 511 keV gamma ray flux to constrain new positron sources sets a conservative bound. A more detailed analysis of positron injection in the galaxy with energies in the $10-10^3$ MeV range would conceivably lead to even stronger constraints~\cite{Beacom:2005qv,Sizun:2006uh}, but it is beyond the scope of this paper.

We focus on positron production in two classes of SN: Type II and Type Ib/c. While both are core-collapse SN, they differ critically in the size of the progenitor star and the density of gas outside the photosphere, both of which play important roles in determining the overall positron flux that escapes the SN. Note however that the collapse of the core and formation of the proto-neutron star are unaffected by the outer layers of the star, hence both Type II and Ib/c SN are well-modeled by the above profile.

The progenitors of Type II SN, specifically the more common Type IIp variety, are thought to be almost exclusively red supergiants (RSGs) ~\cite{Groh:2013mma}. Red supergiants can have stellar radii of over 1400 $R_{\odot}$, or roughly $10^9$ km. Above the star's photosphere, there is a gaseous atmosphere that decreases in molecular number density from a maximum of $\sim10^{10}~\text{cm}^{-3}$ at $10^9$ km in an approximately $r^{-2}$ fashion~\cite{2014MNRAS.437..532M}. Type II SN occur in our galaxy with an average rate of 2 SN/century~\cite{Adams:2013ana}. Therefore, if Type II SN were to contribute $6.3 \times 10^{52}$ positrons per SN, this would saturate the observed galactic positron flux. With this in mind, we take the criterion that any dark photon that allows Type II SN to produce $>10^{53}$ positrons is excluded.

Type Ib/c are similar to Type II SN in that they are core-collapse SN, but they typically occur in hydrogen-stripped stars. Their progenitors are far smaller than RSGs, typically having radii $\lesssim10~R_{\odot}$~\cite{2010ApJ...725..940Y}, and have already shed their outermost layers of hydrogen (and helium, for Type Ic), either due to stripping by a companion star or through stellar winds. Electron densities drop to $10^{12}~\text{cm}^{-3}$ by $\sim 20~R_{\odot}$, or roughly $2\times10^7$ km~\cite{Crowther:2006dd}. Type Ib/c are about an order of magnitude less frequent than Type II SN in our galaxy~\cite{Graur:2016lca}, so for our condition on Type Ib/c SN, we take that any dark photon that allows Type Ib/c SN to produce $>10^{54}$ positrons is excluded.

\subsection{Positron escape}

It is imperative for this bound to apply that the dark photons escape the progenitor star before decaying, otherwise the resulting positrons are likely to annihilate unobserved within the star's outer layers and gaseous envelope. If the positrons are able to escape the star, they have a long lifetime ($10^5$ to $10^6$ years) in the interstellar medium and can slow down and contribute to the observable 511 keV line~\cite{1993ApJ...405..614C}. 

We can estimate the radius $r_{\text{esc}}$ at which an emitted positron would have an order one survival probability using the standard formula for positron annihilation rate in a gas. A conservative choice is to take this to be $\Gamma_{e^+} = \pi r_0^2 n_e$ ~\cite{FRASER196863}, where $r_0$ is the `classical electron radius' ($2.8\times10^{-13}$ cm) and $n_e$ is the electron density, hence we have an interaction length of $\lambda \sim (\pi r_0^2 n_e)^{-1}$. In a Type Ib/c progenitor, $\lambda \sim r$ at roughly 20 $R_{\odot}$, so at radii larger than this, most positrons produced in the decay of the dark photon can escape without annihilation with electrons in the star's outer layers. We therefore set $r_{\text{esc}} = 2\times 10^7$ km for Type Ib/c. By the same argument, we find $r_{\text{esc}} = 10^9$ km for a Type II SN.

Note that since the SN shock begins well within the star, the external layers of the progenitor star do not feel the effects of the SN until the shock has propagated out to them. The shock speed can get up to a third of the speed of light~\cite{refId0}, so in the case of a Type Ib/c SN, it takes roughly $10^2$ seconds for the region at which an emitted positron would escape to feel the effects of the collapse. For this reason, we can treat the outer layers of the star as having no knowledge of the shock when computing the escape probabilities on the timescale of dark photon emission ($\equiv \Delta t\sim$ 10 seconds).

Therefore, for Type Ib/c (Type II) SN we must have the total dark photon flux through a radius $r_{\text{esc}} = 2\times10^7$ km ($10^9$ km) be greater than $10^{54}$ ($10^{53}$) in order to exclude that part of parameter space. The decay width of the dark photon at rest to $e^+ e^-$ in vacuum is given by the following expression:
\be
\Gamma = \frac{1}{3} \alpha_{\text{EM}} \epsilon^2 m' \sqrt{1-\frac{4m_e^2}{m'^2}} \left(1+\frac{2m_e^2}{m'^2}\right)
\ee

It is appropriate to use the vacuum decay width as opposed to the plasma decay width because for the mixing parameters we are considering the dark photons decay length is much larger than the core radius, which is where the finite temperature and density effects are relevant. The decay length is then $d = \frac{\omega v}{m' \Gamma}$ with $\omega$ the dark photon energy, $v$ the velocity, and $\Gamma$ the decay width in its rest frame.

Using this, our constraint for Type Ib/c SN takes the form
\be
\label{eq:conditionIbc}
\Delta t \int dV d\omega \frac{dN}{dV d\omega~dt} e^{-(2\times10^7~\mathrm{km})/d} > 10^{54}~\text{positrons}
\ee
and the equivalent for Type II SN is
\be
\label{eq:conditionII}
\Delta t \int dV d\omega \frac{dN}{dV d\omega~dt} e^{-(10^9~\mathrm{km})/d} > 10^{53}~\text{positrons}.
\ee
Since which of these two conditions provides a stronger bound differs over parameter space, it is more convenient to combine them into a single condition that takes into consideration both contributions from galactic Type II and Type Ib/c SN:
\begin{equation}
\label{eq:condition}
\Delta t \int dV d\omega \frac{dN}{dV d\omega~dt} (e^{-(10^9~\mathrm{km})/d} + 0.1 e^{-(2\times10^7~\mathrm{km})/d} ) > 10^{53}~\text{positrons}.
\end{equation}
This is the constraint we use to generate our bounds.

\subsection{Fireball}

So far we have focused on the effects of the outer layers of the progenitor star on the positron propagation. Another important effect is the interaction of the positrons with the electrons created in the dark photon decay. As pointed out in Ref.~\cite{Kazanas:2014mca}, if a large number of dark photons decay in a shell just outside the photosphere, they can create an optically-thick plasma of electrons, positrons, and photons. Unfortunately, their analysis neglected two important features: the thermal effects on the production of the dark photon (which are large for lighter dark photons) and a detailed analysis of the emission timescale and spectrum of the photons produced, which require more detailed modeling of the plasma. This can be seen clearly, for example, by the fact that the bounds placed in Ref.~\cite{Kazanas:2014mca} are independent of mass at low coupling, which is the result of neglecting thermal effects. As a result, the bounds presented in Ref.~\cite{Kazanas:2014mca} are incorrect and should be replaced by the bounds presented in this paper.

The dynamics of this plasma resembles that of the fireball model~\cite{Piran:1999kx,Meszaros:2006rc} which is used to describe Gamma-Ray Bursts (GRBs). However, it differs from the usual fireball models in an important aspect: the energy density in the plasma is significantly smaller than the one considered for GRBs. To emphasize this difference we will henceforth refer to this plasma as a \textit{dilute} fireball. In these dilute fireballs, both the initial number densities and the temperature are related to the original dark photon number density and mass. Although the dark photons will decay through a large volume, we will focus on a spherical shell of width $d$ and  radius $R_* = \text{min}(d, r_{\text{esc}})$, where $d$ is the average decay length of the dark photons and $r_{\text{esc}}$ is the escape radius discussed in the previous section. An order one fraction of the dark photons that decay outside the star will decay in this shell.

As in the fireball model, it is easiest to understand local properties of the plasma by boosting to a frame that is moving with the plasma, i.e. in which the momentum distribution of the particles in the plasma is isotropic. If the original dark photon flux was monochromatic this frame would coincide with the dark photons' rest frame, and the boost factor would be $\eta = E_{\gprime}/m'$. In our analysis we will neglect the spread in dark photons' initial momenta, which is justified if the velocity spread of the dark photons does not lead to a spatial spread of the dark photon flux that is larger than the average dark photon decay length. Because we are interested in dark photons with lifetimes $\tau \gg 10$ seconds, we can neglect the initial time spread of the dark photons and estimate the initial electron/positron density in the comoving frame as 
\begin{equation}
	n_e(R_*) \sim \frac{\Delta N_\gprime}{4 \pi R_*^2 (d/\eta)}
\end{equation}
where $\Delta N_\gprime$ is the number of dark photons that decay inside the shell and $(d/\eta)$ is the Lorentz-contracted width of the shell. The internal energy of the leptons is determined by the mass of the dark photon, and therefore if they form a plasma, the initial temperature is $T\sim m'$.

One can easily check that the initial number density of the plasma is significantly below the thermal number density of a plasma with the same temperature ($\sim m'^3$). This leads to a much smaller optical depth than in the typical fireball model. The initial optical depth for the shell is well-approximated as
\begin{equation}
	\tau_{0} \approx n_e \sigma_{e^+ e^- \rightarrow \gamma \gamma} (d/\eta) \sim \left(\frac{\Delta N_\gprime}{4 \pi R_*^2 (d/\eta)}\right)\left(4\pi\frac{\alpha^2}{T^2}\right)(d/\eta) \sim \left(\frac{\alpha^2~ \Delta N_\gprime}{m'^2 R_*^2}\right)
\end{equation}
If it is larger than one, a plasma forms and the reaction $e^+ e^- \leftrightarrow \gamma \gamma$ is in local equilibrium and the photon and lepton number densities are related by detailed balance.\footnote{In the presence of large magnetic fields the optical depth could be enhanced due to charged particles being trapped by the fields. However one can easily show that the kinetic energy in the plasma is orders of magnitude larger than the expected energy density in magnetic fields in the region of interest. Therefore, the magnetic field would be combed out by the plasma and would not lead to efficient trapping. (For details on radially-combed magnetic fields, see e.g.~\cite{1993ApJ...405..614C}.)}

The initial temperature of the plasma is at least as large as the dark photon mass, and hence initially the total number densities of positrons and photons are comparable. The only way to have a significant decrease in the number of positrons is if there is an efficient way to cool the plasma so that the temperature becomes smaller than the electron mass $m_e$. There are two important ways in which the plasma can cool: internal processes and adiabatic expansion.\footnote{Large baryon loading could also in principle lead to significant cooling. However, since we are considering decays in regions where the stellar atmosphere is very diffuse, one can show that the number density of baryons is too small to lead to appreciable cooling.}

Internal processes are important if reactions that can change the total number of particles, in particular 2-to-3 reactions (e.g. $e^+ e^- \rightarrow 3 \gamma$), are occurring in the plasma. Such processes increase the number density, driving it towards $T^3$, and consequently lower the temperature due to energy conservation. The optical depth for such reactions is approximately a factor of $\alpha$ lower than the optical depth for the 2-to-2 process. In Fig.~\ref{Fig: bounds}, we indicate the region in which the 2-to-3 optical depth is larger than 0.1 with a red dashed line. In this region, we expect that a large fraction of positrons annihilate and therefore cannot place a positron injection bound there. However, we can place a bound on this region using the resulting gamma rays (see Section IV.B).

Even if 2-to-3 processes are off, if 2-to-2 processes are on, the plasma can still cool due to adiabatic  expansion, as occurs in a thin region of our parameter space. Unlike in the situation when 2-to-3 processes are on, we can still place a positron injection bound in this region. The cooling due to the expansion is analogous to the case of the regular fireball~(see Ref.~\cite{Piran:1999kx,Meszaros:2006rc} for a detailed discussion). As the plasma expands it converts the internal gas energy (temperature) to kinetic energy of the expanding shell, i.e. as the plasma expands, most of the momentum becomes radial. For a plasma with energy density dominated by radiation, this leads to the scaling $T \propto 1/R$, where $R$ is the radius of the expanding shell in the original star frame. As a result, when $T>m_e$, the cross-section scales as $\sigma \propto 1/E_{\text{CM}}^2 \propto 1/T^2 \propto R^2$ and the number density  scales as $n_e(R) \propto 1/R^2$ (where we ignore the growth of the shell width as it cancels in the optical depth). Therefore, one can easily see that the optical depth remains constant while $T > m_e$.

Eventually, at some radius $R_e$, the temperature will drop below the electron mass, causing the number density and cross-section to begin scaling differently. The number density will scale as
\begin{equation}
\label{eq:positron-suppression}
	n_e(R) \approx n_e(R_e)\left[ \left(\frac{R_e}{R}\right)^2 \left(\frac{m_e}{T}\right)^{3/2} e^{-m_e/T}\right]
\end{equation}
due to Boltzmann suppression and the cross-section begins to scale as $\sigma \propto 1/(E_{\text{CM}}^2 v)  \propto T^{-1/2} \propto R^{1/2}$ since $E_{\text{CM}}$ is fixed at roughly $m_e$ but the velocity $v \sim \sqrt{T/m_e}$. Therefore, we see immediately that the $R$-dependence cancels in the optical depth except for an exponential factor and $\tau$ scales as $\tau \propto \exp(-R/R_e)$.

In the regions where 2-to-2 reactions are occurring but 2-to-3 are not, we necessarily have that the 2-to-2 optical depth is $1 \lesssim \tau \lesssim 10$. Therefore, to decrease this optical depth below one such that positrons can escape, we see that we must only expand by at most $\sim 3 R_e$. Using this and Eq.~\ref{eq:positron-suppression}, we see that the total number of positrons escaping the supernova is only an order-one factor smaller than the initial number produced at $R_*$, hence the formation of a fireball has little ultimate effect on the flux when it is sufficiently dilute such that number-changing processes are not occurring. We include this small suppression when computing the bounds in Fig.~\ref{Fig: bounds}.

\section{Gamma-ray Bounds}

In this section, we introduce two distinct gamma-ray signatures of SN-produced dark photons that we then use to constrain new regions of dark photon parameter space.

\subsection{SN 1987a}

There is an additional signal that can be used to constrain dark photons in this parameter space, namely the absence of an observed gamma-ray flux above background in the first few minutes following the arrival of SN1987a's neutrinos on Earth.

The non-observation of an increased gamma-ray flux by the Gamma Ray Spectrometer (GRS) aboard the Solar Maximum Mission results in the following fluence limits for the first 270 seconds, taken from Ref.~\cite{PhysRevLett.62.509}: $f_{\gamma} \lesssim 5~\cm^{-2}$ (4-6 MeV); $f_{\gamma} \lesssim 1.6~\cm^{-2}$ (10-25 MeV); and $f_{\gamma} \lesssim 3.3~\cm^{-2}$ (25-100 MeV). Since the exact spectrum of the outgoing gamma rays depends on the choice of profile, we choose to place a conservative limit by summing the fluences in all bins and requiring that the total number of gamma rays produced does not exceed it. This results in a fluence limit of $f_{\gamma} \lesssim 10~\cm^{-2}$, which translates to a limit on the number of hard gamma rays escaping the supernova of $N_{\gamma} < 4\times10^{48}$.

In order to use the quoted fluence limit, we must require that the arrival time of the resulting gamma rays on Earth be less than 270 seconds after the arrival time of the neutrinos. The massive dark photons are not traveling exactly at the speed of light, so before they decay, they accumulate some time delay behind the neutrinos. We therefore compute a minimum boost required such that the dark photons reach the distance at which the gamma rays are primarily produced before this delay reaches 270 seconds. This distance is taken to be $R_* = \text{max}(r_{\text{1987}}, d)$, with $d$ the decay length of the dark photon and $r_{\text{1987}}$ the escape radius for SN1987a's progenitor ($r_{\text{1987}} = 4\times10^7$ km~\cite{1538-3873-104-679-717}) since if $d > r_{1987}$, then the majority of decays will occur by $r=d$.

We consider this bound only in the regime where a plasma never forms since, as argued above, it is likely that a fireball would result in gamma rays below the GRS lower limit of 4 MeV. In this region of parameter space, hard gamma rays can be produced as final state radiation with a rate that is suppressed by roughly $\alpha$ with respect to the exclusive $e^+ e^-$ decays.
Therefore the number of these hard gamma rays produced is
\be
(\Delta t) \int dV \int_{\gamma_{\text{min}}}^\infty d\omega \frac{dN}{dV dt~d\omega}\left( e^{-r_{1987}/d} \text{Br}(e^+ e^- \gamma_{> 4 \mev} ) \right)
\ee
where $\Delta t = 10$ seconds is the emission timescale, $\gamma_{\text{min}}$ is the minimum boost required such that the decays occur within 270 seconds, and $ \text{Br}(e^+ e^- \gamma_{> 4 \mev} )$ is the branching ratio for a dark photon with energy $\omega$ to decay to a photon with energy above $4$ MeV. (See Appendix B for a derivation of this branching ratio.) If this quantity ever exceeds $4\times10^{48}$ gamma rays, we place a constraint on the corresponding region of parameter space. This effect also has discovery potential: if we were to measure the spectrum of the next galactic SN and find an anomalous gamma-ray excess, it could provide evidence for dark photons. (See Section V.)

Ref.~\cite{Kazanas:2014mca} attempts to use the same GRS non-observation to place bounds on dark photons using gamma rays produced in a fireball, but neglects to compute the outgoing spectrum of the resultant gamma rays, which are expected to have energies $< 4$ MeV. There does not appear to be any data for these lower energy bins, hence a bound cannot be placed by GRS in this regime.

\subsection{Diffuse extragalactic flux}

The only remaining region of parameter space left to constrain is the region in which a fireball forms and 2-to-3 processes are occurring. This can be partially accomplished by comparing the diffuse extragalactic flux of gamma-rays that would be generated by these fireballs to the measured value.

It is well-known that extragalactic supernovae lead to a diffuse background of supernova neutrinos. In the same fashion, the collective effect of extragalactic supernovae producing these 2-to-3 fireballs would be to contribute to the diffuse extragalactic background of gamma rays. We can easily compute the diffuse energy flux in a given energy bin on Earth by performing a line-of-sight integral over distance in the following way~\cite{Beacom:2010kk}:
\be
\label{eq: snrate}
\Phi_{\gamma} = \int_0^{\infty} E_{\gamma}^{\mathrm{bin}} R_{SN}(z)\left|\frac{dt}{dz}\right|~dz
\ee
where $E_{\gamma}^{\mathrm{bin}}$ is the total energy of gamma-rays in the specified bin produced by a single SN, $R_{SN}(z)$ is the redshift-dependent supernova rate (taken from~\cite{Beacom:2010kk}), and $(\frac{dt}{dz})^{-1} = H_0 (1+z) \sqrt{ \Omega_{\Lambda} + \Omega_{m} (1+z)^3}$.

We restrict our attention to gamma rays with energy between 100 keV and 4 MeV. This is simply due to the fact that for energy greater than 4 MeV, the GRS bound would apply and by $T\sim100$ keV,  the fireball has become transparent to gamma-rays. The extragalactic flux in this bin has been measured by the Solar Maximum Mission (SMM)~\cite{doi:10.1063/1.1303252} and the High Energy Astronomy Observatory (HEAO-1)~\cite{0004-637X-475-1-361}, which allows us to integrate over this bin to find a total flux on Earth in this energy range. Performing the integral, we get a value of roughly $0.5~\Mev~\cm^{-2}~ \mathrm{s}^{-1}$. It is therefore simple to determine what total energy released by a single supernova would lead to an excess over the measured extragalactic flux. This calculation yields a bound of $\sim 1.2\times10^{56}$ MeV in gamma rays with $0.1~\Mev<E_{\gamma}<4~\Mev$ for Type II SN and $\sim 1.2\times10^{57}$ MeV for a Type Ib/c. These bounds are conservative in that they assume the spectrum of gamma rays has the same shape as the background. A more careful analysis of the spectrum would improve these bounds.

The total initial energy of the dark photons that escape the star has to be released from the supernova in some form. If even a small fraction of it were released as positrons, we would be able to place a positron bound on the 2-to-3 fireball region. Hence we assume here that all the energy is released in the form of gamma rays, which allows us to simply compare the total energy of the dark photons that decay outside of $r_{\text{esc}} = 2\times10^7$ km to the extragalactic bound on energy we have just computed.\footnote{We focus purely on Type Ib/c SN since in this region of parameter space, the dark photon decay length is significantly less than the radius of a Type II progenitor, hence the escaping flux from a Type II SN suffers from a large exponential suppression.} We find that this simple argument covers a band of parameter space within the 2-to-3 fireball region (``Gamma rays (diffuse flux)'' in Fig. \ref{Fig: bounds}), though does not constrain the entirety of the region. Regions in which this fails to place a constraint are colored in yellow. A careful analysis of the spectrum of gamma rays emitted by the expanding fireball would likely constrain these regions as well, but it is beyond the scope of this paper.

\section{Discovery potential}

In this section, we explain how a future observation of the next galactic SN could allow us to discover the dark photon through gamma rays produced in its decays.

While the previous sections have addressed bounds we can place on dark photons using existing observations, we can additionally consider the case of a future nearby supernova and ask in what regions of parameter space we would have sensitivity to discovering the dark photon. With a galactic SN rate of $\sim 2$ per century, it is not unreasonable to expect to observe one in the next few decades. With this in mind, we seek to compute a rough estimate of the parameter space we could probe with such a future observation.

The signature we will be searching for is the same as for the SN1987a bound, namely the gamma-ray flux from hard photons produced in three-body decays of the dark photon. We take as our background a conservative estimate of the diffuse gamma-ray flux from the galactic center in the energy band 1-100 MeV, as this exceeds the expected gamma-ray flux from the SN itself in the absence of BSM processes. As measured by COMPTEL~\cite{1996A&AS..120C.381S}, a gamma-ray telescope aboard the NASA Gamma Ray Observatory, the background for this bin can be estimated as $E^2 \frac{d\Phi}{dE} \lesssim 0.01~\Mev~\cm^{-2}~ \text{s}^{-1}~ \text{sr}^{-1}$. Taking $d = 8$ kpc (the distance from Earth to the center of the galaxy), this becomes a rate of $(dN/dt)_{\gamma} \sim 10^{44}~\text{s}^{-1}$. Note that this is not a true production rate of gamma rays by the SN, but is rather the total number of gamma rays ultimately produced divided by the time interval over which they arrive at Earth.

In order to determine the dark photon production required to exceed this gamma-ray flux, consider a shell of dark photons emitted from the SN. As dark photons decay, the associated gamma rays escape at $c$ while the remaining dark photons lag behind. Once an $\mathcal{O}(1)$ fraction of the dark photons have decayed (after roughly $\gamma_{\text{avg}}\tau$, their Lorentz-dilated lifetime), the gamma-rays will have a time-spread of roughly $(1-v)\gamma_{\text{avg}}\tau$, where $v$ is the average velocity of the dark photons and $\gamma_{\text{avg}}$ is the average boost of the dark photons. Therefore, to find a decay rate, we simply divide the total dark photon production by this time-spread (and multiply by the fraction of decays that will produce a gamma ray with energy $> 1$ MeV, $\text{Br}(e^+ e^- \gamma_{> 1 \mev} )$, the derivation of which appears in Appendix B). Hence we have the relation
\be
(dN/dt)_{\gamma} = \text{Br}(e^+ e^- \gamma_{> 1 \mev} )N_{A'}~\text{min}\left[ \left(\frac{1}{1-v} \right)\frac{1}{\gamma_{\text{avg}}\tau},~\frac{1}{10~\sec}\right]
\ee
The minimum function appears only because the time-spread can never be smaller than the initial 10-second emission timescale of the dark photons. Note that $N_{A'}$ is the total number of dark photons that will escape the SN and is simply given by the expression on the left-hand side of Eq.~\ref{eq:conditionIbc} for Type Ib/c and Eq.~\ref{eq:conditionII} for Type II. Using our condition that $(dN/dt)_{\gamma} > 10^{44}~\text{s}^{-1}$, we can estimate the sensitivity limits.

Since it is roughly an order of magnitude more likely that the next supernova we observe will be a Type II SN, we have indicated those bounds in solid blue in Figure \ref{Fig: bounds}. If it were to be a Type Ib/c, however, the smaller escape radius would allow us to be sensitive to an even larger region of parameter space, which we have chosen to indicate with a dotted blue line in Figure \ref{Fig: bounds}. Recall that an assumption used to compute these bounds is that no fireball forms, hence they do not extend into the unconstrained yellow regions in Figure \ref{Fig: bounds}. However, observations of the $\sim100$ keV gammas produced in the fireball of a nearby SN would likely constrain these regions as well.

Though it may seem surprising that the Type II discovery region does not enclose the SN1987a gamma-ray bound, this is due entirely to the fact that SN1987a's escape radius was $\sim 4 \times 10^{7}$ km, which is two orders of magnitude smaller than the conservative $10^9$ km we use for a generic Type II progenitor. It is clear that the Type Ib/c discovery region (with $r_{\text{esc}} = 2\times 10^7$ km) fully encloses the SN1987a bound.

\section{Results}

\begin{figure}
  \centering
  \includegraphics[width=0.8\textwidth]{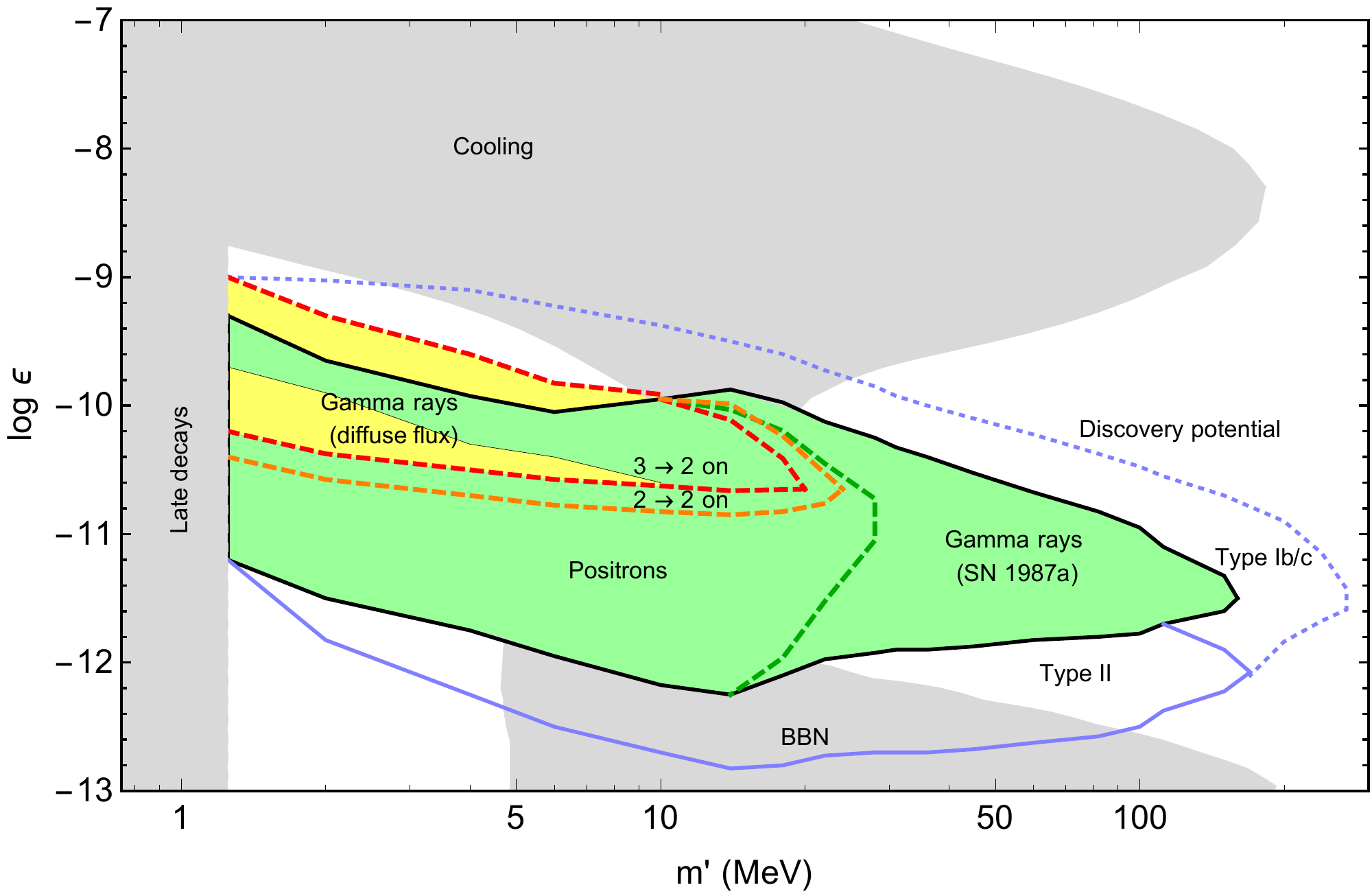}
  \caption{The overall region excluded by this work is colored in green. The blue curves show the region in which we could discover the dark photon during the next galactic SN. The solid blue curve is drawn considering the case of a Type II SN, the dotted blue curve a Type Ib/c. Yellow regions indicate where a simple analysis of the diffuse flux fails to place a constraint, though a more involved analysis would likely constrain these regions as well. The positron bound rules out parameter space between the red and the green dashed lines. The gamma-ray constraint from SN1987a extends the excluded region to higher masses (past the green dashed line), and the diffuse gamma-ray flux to stronger couplings (above the red dashed line). The orange dashed line indicates where a plasma forms but only 2-to-2 processes are on, whereas the red dashed line indicates where a fireball with 2-to-3 processes occurring forms. Previous bounds are shown in gray. The cooling bound is reproduced from~\cite{Chang:2016ntp} and the BBN constraint from~\cite{Fradette:2014sza}. The late decay bound is taken from~\cite{Redondo:2008ec} and comes from considering the decay of dark photons on cosmological timescales (see also~\cite{Essig:2013goa}). 
    \label{Fig: bounds}}
\end{figure}

For any given $m'$ in our range of interest, we can easily invert our criteria to determine at what $\epsilon$ the positron or gamma-ray flux becomes too large. This places bounds on regions of parameter space up to two orders of magnitude below current bounds on dark photons from cooling constraints. The result is shown in Figure \ref{Fig: bounds}. The bound is cutoff on the lower end of $m'$ by the fact that the dark photon must have $m' > 2m_e$ in order for the decay to $e^+ e^-$ to take place.

The upper bound corresponds to where the diminishing decay length ($d \propto \epsilon^{-2}$) results in too few dark photons escaping out of the star before decay (for larger $\epsilon$ it is possible that the energy carried by the dark photon and deposited at the outer layers of the star would lead to an early explosion which could serve as a further constraint in this region of parameter space~\cite{Kazanas:2014mca}). The lower bound corresponds to where the mixing becomes too small to produce enough dark photons to violate our conditions. Note that contributions to the lower positron bound are dominated by Type II SN since at mixings this low, the exponential suppression due to decay length is not a large effect and Type II SN are an order of magnitude more common than Type Ib/c. The upper positron bound is dominated by contributions from Type Ib/c since at these higher couplings, the decay length becomes much shorter so a smaller escape radius compensates for the order of magnitude loss in abundance in comparison to Type II. The 3-body gamma-ray bound at higher masses uses SN1987a throughout. This gamma-ray bound extends to higher masses than the positron bound because it requires several orders of magnitude fewer decay products than does the positron bound. Despite this, it does not place stronger constraints on low $\epsilon$ than the positrons below 20 MeV due to the increasing decay lengths resulting in fewer decays within the 270-second window.

The region within the red dashed line denoted ``Gamma rays (diffuse flux)'' is the region in which number-changing processes may be occurring in the resulting electron-positron plasma outside the photosphere. Here, we use our bound from the diffuse extragalactic gamma-ray flux. The yellow regions indicate where a number-changing fireball forms but the diffuse bound fails to constrain the parameter space. The fireball no longer forms above the upper red dashed line due to too few dark photons escaping. Note that the sharp features in the upper yellow region are a numerical artifact from computing with a finite grid in parameter space and are not physical. Between the orange and red dashed lines is where only 2-to-2 process are occurring, so we have a plasma but can still place a positron injection bound (see Section III.C). Positron bounds end at the green dashed line, beyond which we can constrain parameter space using our bound from observations of SN1987a.

The blue curves in the figure indicate regions in which we would be able to observe the gamma-ray flux associated with dark photon decays if a supernova were to occur near the galactic center. The solid blue line shows the region for a Type II SN while the dotted blue line shows the region for a Type Ib/c. The Type II region does reach to as strong a coupling as the Type Ib/c region because of the exponential suppression of the escaping flux due to the dramatically larger escape radius for Type II progenitors. However, since Type II SN are an order of magnitude more abundant than Type Ib/c, it is unlikely that the next galactic supernova will be a Type Ib/c. The curve is drawn simply for completeness.

\section{Conclusion}

We have used direct signals to place new bounds on models of kinetically-mixed dark photons by considering various effects of the electromagnetic decays of dark photons produced in supernovae. We have shown that if any novel particle produced within a Type II supernova produced in excess of $10^{53}$ escaping positrons (or $10^{54}$ in a Type Ib/c), it would result in a 511 keV galactic emission in conflict with current observations by SPI/INTEGRAL. Additionally, we used the result that dark photons could not lead to an emission of greater than $4\times10^{48}$ gamma rays with $E_{\gamma} > 4$ MeV in the 270-second window following the arrival of SN1987a's neutrinos without being excluded by observations by GRS. Finally, we used SMM and HEAO-1's observations of the diffuse extragalactic gamma-ray flux to constrain the dark photons produced in any given Type Ib/c supernova from releasing more than $\sim 1.2\times10^{57}$ MeV in gamma rays with $0.1~\Mev<E_{\gamma}<4~\Mev$. By combining these constraints, we are able to exclude a large region of parameter space two orders of magnitude weaker in coupling than current cooling bounds. Excitingly, we have also shown the potential for a discovery of the dark photon by observing the gamma ray spectrum of the next galactic supernova. These conclusions apply to any model in which the dark photon has an $\mathcal{O}(1)$ branching ratio to $e^+ e^-$. Our results are summarized in Fig. \ref{Fig: bounds}.

\section*{Acknowledgements}

The authors thank Ed Hardy, Robert Lasenby, and Sam McDermott for useful discussions.
W.D., G.M.T., and P.W.G. are supported by DOE Grant DE-SC0012012. W.D. and P.W.G are further supported by NSF Grant PHY-1720397, the Heising-Simons Foundation Grants 2015-037 and 2018-0765, DOE HEP QuantISED award \#100495, and the Gordon and Betty Moore Foundation Grant GBMF7946. The work of G.M.T. was also supported by the NSF Grant PHY-1620074 and by the Maryland Center for Fundamental Physics. S.R. is supported in part by the NSF under grants PHY-1638509 and PHY-1507160, the Simons Foundation Award 378243, and the Heising-Simons Foundation Grant 2015-038. D.K. is supported in part by the U.S. Department of Energy, Office of Science, Office of Nuclear Physics, under contract number DE-AC02-05CH11231 and DE-SC0017616, by a SciDAC award DE-SC0018297, by the National Science Foundation Grant PHY-1630782, and by the Heising-Simons Foundation Grant 2017-228.

\begin{appendix}

\section{Profile dependence}

In order to estimate the uncertainty on our results due to our choice of SN profile, we compared to a profile with significant differences in both dependence and overall magnitude. The profile chosen as a comparison is given by the following.

\be
\rho(r) = \rho_0 \times
\begin{cases}
e^{-2(r-R_0)/R_0} & r < R_0 \\
e^{(R_0 - r)/h} & R_0 \leq r < R_t \\
e^{(R_0 - R_t)/h}  (r/R_t)^{-3} & r \geq R_t 
\end{cases}
\ee
\be
T(r) = 
\begin{cases}
T_{\text{in}}+(T_0 \frac{R_0}{R_{\text{in}}} - T_{\text{in}})\exp\left[-16\frac{(r-R_{\text{in}})^2}{R_{\text{in}}^2}\right] & r < R_{\text{in}} \\
T_0 \left(\frac{R_0}{r}\right) & R_{\text{in}} \leq r < R_0 \\
T_0 e^{(R_0 -r)/4h} & R_0 \leq r < R_{\nu} \\
T_0 e^{(R_0 -r)/4h} (R_{\nu}/r) & r \geq R_{\nu}
\end{cases}
\ee
\be
Y(r) = 
\begin{cases}
Y_{\text{in}}+(Y_0 - Y_{\text{in}})\exp\left[-16\frac{(r-R_{\text{in}})^2}{R_{\text{in}}^2}\right] & r < R_{\text{in}} \\
Y_0 + (Y_t  - Y_0) \exp\left[-100\frac{(r-R_{\text{in}})^2}{R_{\text{in}}^2}\right] & R_{\text{in}} \leq r < R_t \\
Y_t + (Y_{\text{out}}  - Y_t) \frac{r-R_t}{R_{\text{out}} - R_t} & R_t \leq r < R_{\text{out}} \\
Y_{\text{out}} & r \geq R_{\text{out}}
\end{cases}
\ee

We used the following parameters:
\begin{align*}
R_{\text{in}} &= 8~\km \\
T_{\text{in}} &= 15~\Mev \\
Y_{\text{in}} &= 0.25 \\
R_0 &= 15~\km \\
\rho_0 &= 10^{14}~\text{g cm}^{-3} \\
T_0 &= 27.5~\Mev \\
Y_0 &= 0.1 \\
R_{\nu} &= 21~\km \\
R_t &= 25~\km \\
h &= 1~\km \\
Y_t &= 0.4 \\
R_{\text{out}} &= 30~\km \\
Y_{\text{out}} &= 0.5
\end{align*}

\begin{figure}
  \centering
  \includegraphics[width=0.9\textwidth]{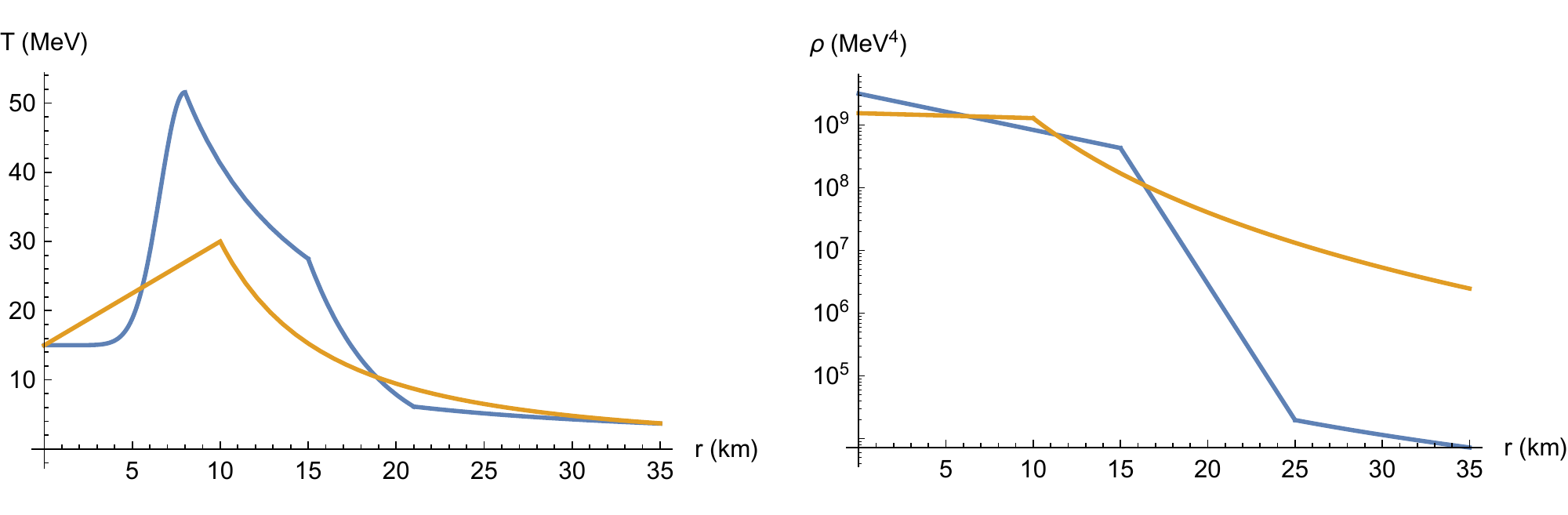}
  \caption{The two SN profiles compared in order to assess the effect of modeling uncertainties on our results. The yellow curve is the profile used to generate our bounds while the blue curve is the one described in this appendix. Though the blue curve reaches a much higher temperature than the yellow curve, the overall effect on our bounds is small. Note that the disagreement in densities at radii outside 20 km does not cause a significant discrepancy in our results because the majority of production occurs in the dense regions within the core.
    \label{Fig: profiles}}
\end{figure}

The parameters were chosen such that the profile coincided well with simulation results produced by DK. The temperature was increased to a maximal realistic value (peak temperature of roughly 50 MeV) in order to exaggerate any differences with the cooler profile used in the body of this paper. It was found that even this dramatic change in profile  increased the bounds by roughly $\Delta(\log{\epsilon}) \sim 0.1$ and everywhere, the change was $\Delta(\log{\epsilon}) \leq 0.3$. We conclude that the results presented in this paper are not highly sensitive to the choice of SN profile.

\section{Dark photon decays with hard photon emission}
\label{section:branching-photon}

Here we describe the calculation of the branching ratio of a dark photon with boost factor $\eta$ to $e^+e^- \gamma$ with a final state photon energy above some $\omega_{\text{min}}$. At tree level, the averaged matrix element for the decay $A^\prime \rightarrow e^+e^- \gamma$ is
\begin{equation}
	\begin{aligned}
	 \bar{|\mathcal{M}|^2} = & \frac{8 \epsilon^2 e^4}{3(m_{23}^2-m_e^2)^2 (m'^2+m_e^2-m_{12}^2-m_{23}^2)^2} \\
	 &\left[ m'^6(m_{23}^2-3 m_e^2)-m'^4 \left(3 m_{23}^4 -6 m_{23}^2 m_e^2 + m_{12}^2(m_{23}^2-5 m_e^2) + 7 m_e^4 \right) \right. \\
	 & + m'^2 (m_{12}^4(m_{23}^2-3 m_e^2)+4(m_{23}^2-m_e^2)^3+4 m_{12}^2(m_{23}^4-m_{23}^2 m_e^2+2 m_e^4)) \\
	 &-m_{12}^6(m_{23}^2-m_e^2)-2(m_{23}^2-m_e^2)^4 \\
	 &\left. - 4 m_{12}^2 m_{23}^2 (m_{23}^2-m_e^2)^2 - m_{12}^4 (3 m_{23}^4 - 2 m_{23}^2 m_e^2 + 3 m_e^4) \right] ,
	\end{aligned}
\end{equation}
where $m_{12}^2 = (k+q)^2$ and $m_{23}^2=(k+l)^2$ with $k$, $q$, and $l$ the electron, positron, and photon momenta respectively.

In the dark photon rest frame, the differential partial width of this decay is given by
\begin{equation}
	\label{eq:branching-gamma}
	\frac{d^2 \Gamma}{d E_\gamma~d \cos \theta} = \frac{1}{(2 \pi)^3~32 m'^2} \int dm_{23}^2 ~\bar{|\mathcal{M}|^2},
\end{equation}
with $E_\gamma$ the energy of the final-state photon and $\theta$ the angle between the photon momentum and the $\hat z$-direction. The integration limits are
\begin{equation}
(m_{23}^2)_\text{max/min} = (E_2^* + E_3^*)^2- \left( \sqrt{E_2^{* 2}-m_e^2}  \mp E_3^* \right)^2 \, ,
\end{equation}
with
\begin{equation}
	E_2^* = \frac{m_{12}}{2} \ \ \ \text{and} \ \ \ E_3^* = \frac{m'^2-m_{12}^2}{2 m_{12}} \, .
\end{equation}

If we require that the energy of the photon is above $\omega_{\text{min}}$ in the frame in which the dark photon has a boost $\eta$ in the $\hat z$-direction, this translates to a minimum energy
\begin{equation}
	(E_\gamma)_{\text{min}} = \frac{ \omega_{\text{min}}}{\eta + \cos \theta \sqrt{\eta^2-1}}
\end{equation}
in the rest frame. Using this constraint, we can integrate Eq.~\ref{eq:branching-gamma} over $E_{\gamma}$ and $\theta$ to obtain the partial width of the boosted dark photon with a final state photon with energy above $\omega_{\text{min}}$.

\section{Axion extragalactic gamma-ray background}

Since the original publication of this paper, the authors have received many queries about whether or not a bound could be placed on the diffuse extragalactic flux of gamma-rays that axion decays would produce. The answer is that the bounds are very weak and only in regions of parameter space already disfavored by cosmology. However, we have chosen to include an appendix that addresses this directly such that the calculation and result are in the literature for the community to reference. Note that while we use the term ``axion'' throughout, the particles in question are not the QCD axion but rather generic axion-like particles.

MeV-scale axions have a decay mode to two photons with width $\Gamma_{a\gamma\gamma} = \frac{g_{a\gamma\gamma}^2 m_a^3}{64 \pi}$ ~\cite{Jaeckel:2017tud}. 
As a result, the SN axions may escape the SN, then decay to gamma rays. The collective effect of all extragalactic SNe would then be a diffuse flux of gamma rays produced by axions. This would form a contribution to the extragalactic gamma-ray background which would be ruled out if it exceeds the measured flux.

We considered two cases for the axions: one in which they couple only to photons and one in which there is a coupling to nucleons of the same strength ($g_{aNN} = g_{a\gamma\gamma}$). To compute the production solely due to Primakoff emission, we replicated the results of Section 2 of Ref.~\cite{Payez:2014xsa}, which takes into account nuclear degeneracy effects in the SN core. To compute the production of axions with a coupling to nucleons, we followed the methodology described in Sections 4 and 5 of Ref.~\cite{Chang:2018rso}, which includes corrections to the bremsstrahlung rate that were overlooked in earlier literature.

We chose to use two fairly different analytic SN profiles in order to determine the sensitivity of the result on the choice of profile. The first is the fiducial profile used in Ref.~\cite{Chang:2016ntp} (noted as ``Cooler profile'' on the plot), the second is the profile used in the present paper (noted as ``Hotter profile'' on the plot).

Specifying the production rate and profile determines the spectrum of outgoing axions and by extension, gamma-rays, which will have roughly half the energy of the axions. Unlike the case of the dark photon, in the regions of parameter space of interest, the axion decay length is much larger than the radius of the progenitor star, hence all axions escape before decay. However, the decay length is actually so long that towards lower masses, we must include an exponential suppression for when the lifetime exceeds the light-travel time to the average SN ($z\sim 0.8$).

We carry out the computation for the EGRB produced by the decays of axions via Eq.~\ref{eq: snrate} of this work. Note that since the average SN occurs at $z\sim 0.8$, the EGRB spectrum will be redshifted by a factor of $\sim 2$ in comparison to the local spectrum computed for a single SN. 

To place a bound, we compare the predicted axion contribution to the EGRB in each energy bin measured by COMPTEL~\cite{doi:10.1063/1.1307028} and to the 2$\sigma$ upper limit on the EGRB set by the COMPTEL measurement in that bin. An axion that produces an EGRB contribution that exceeds the COMPTEL measurement in any bin is considered ruled out.

\begin{figure}
  \centering
  \includegraphics[width=0.95\textwidth]{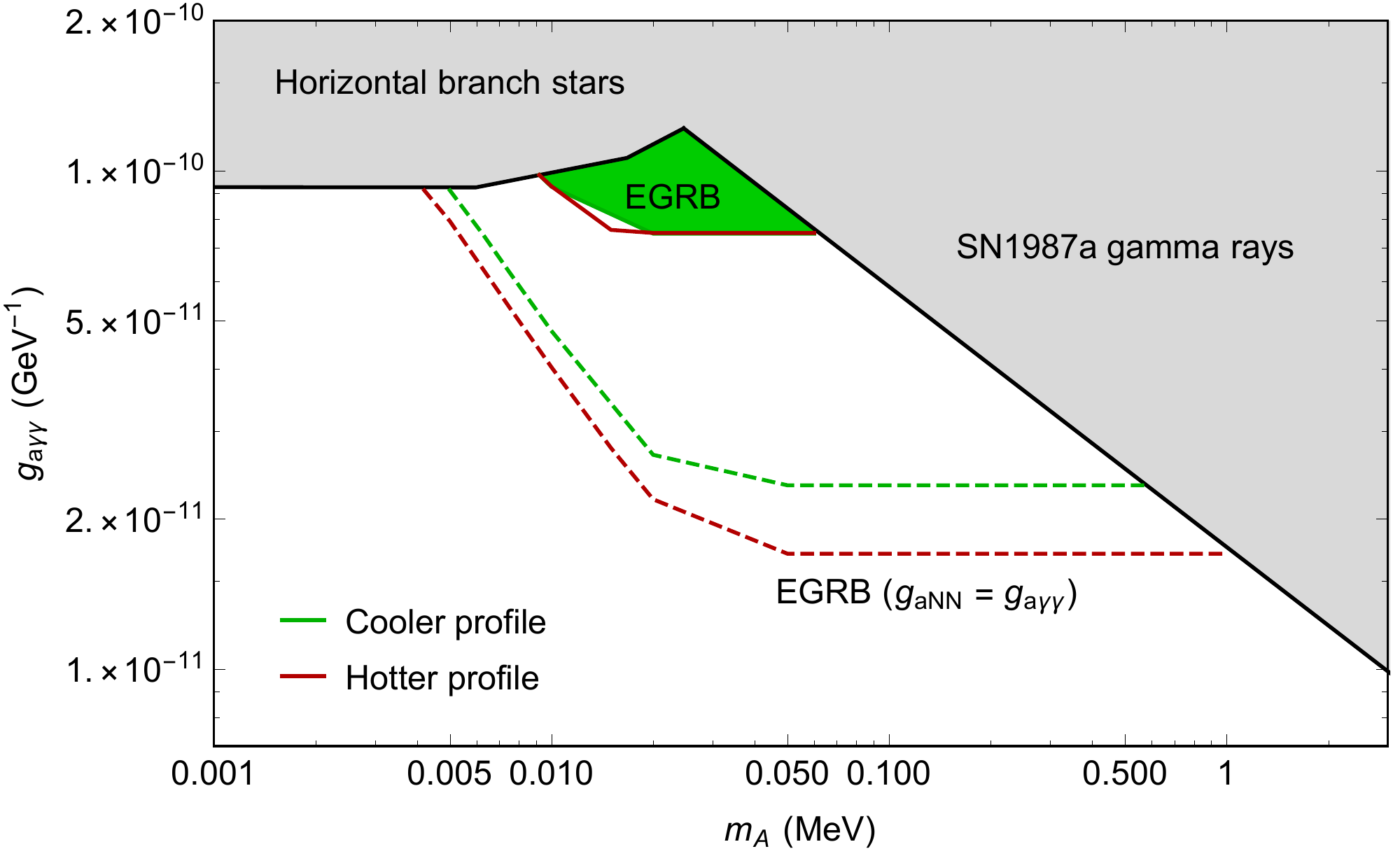}
  \caption{Bounds on the axion from the EGRB. Solid lines denote bounds computed using exclusively Primakoff production with no axion coupling to nucleons, dashed lines denoted bounds computed with the inclusion of nucleon bremsstrahlung as a source (with $g_{aNN} = g_{a\gamma\gamma}$). The green bounds correspond to the fiducial profile of Ref.~\cite{Chang:2016ntp} while the red bounds correspond to the profile in this work. The most conservative possible bound one could set would be the shaded green region, which corresponds to the colder profile and no coupling to nucleons. The resulting bounds are within the uncertainties of existing bounds.  Note further that the new bounds lie entirely within existing cosmological bounds, though these can be evaded with different reheating histories.
     \label{Fig: bound}}
\end{figure}

The resulting bounds are shown in Figure~\ref{Fig: bound}. Very little new parameter space is constrained. Since the lifetime depends so steeply on the mass, walking towards lower masses takes us rapidly into a regime where the axion lifetime becomes longer than the age of the universe, shutting off the bound. The differences between the two profiles are to be expected in that they have different core temperatures. The profile used for the red bounds is slightly hotter than the profile used for the green bounds, hence results in a slightly larger flux of axions. Furthermore, these bounds lie entirely within a region of parameter space in which axions can dramatically affect cosmology, though the associated bounds can be evaded with different reheating histories~\cite{Jaeckel:2017tud}. Finally, it should be noted that the bounds in which we have turned on a nucleon coupling equal to the photon coupling are a bit misleading in that the adjacent SN1987a bounds were computed purely with Primakoff production and would eat into this parameter space if bremsstrahlung had been included. 

As appealing of a notion as it is to constrain axion decay with the EGRB, it is not as constraining as one might hope.

\end{appendix}

\bibliographystyle{JHEP}
\bibliography{ref}

\providecommand{\href}[2]{#2}\begingroup\raggedright\begin{thebibliography}{10}

\bibitem{Essig:2013lka}
R.~Essig et~al., \emph{{Working Group Report: New Light Weakly Coupled
  Particles}},  in \emph{{Proceedings, 2013 Community Summer Study on the
  Future of U.S. Particle Physics: Snowmass on the Mississippi (CSS2013):
  Minneapolis, MN, USA, July 29-August 6, 2013}}, 2013,
  \href{https://arxiv.org/abs/1311.0029}{{\ttfamily 1311.0029}},
  \href{http://www.slac.stanford.edu/econf/C1307292/docs/IntensityFrontier/NewLight-17.pdf}{http://www.slac.stanford.edu/econf/C1307292/docs/IntensityFrontier/NewLight-17.pdf}.

\bibitem{Alexander:2016aln}
J.~Alexander et~al., \emph{{Dark Sectors 2016 Workshop: Community Report}},
  2016, \href{https://arxiv.org/abs/1608.08632}{{\ttfamily 1608.08632}},
  \href{http://lss.fnal.gov/archive/2016/conf/fermilab-conf-16-421.pdf}{http://lss.fnal.gov/archive/2016/conf/fermilab-conf-16-421.pdf}.

\bibitem{Raffelt:1996wa}
G.~G. Raffelt, \emph{{Stars as laboratories for fundamental physics}}. 1996.

\bibitem{Holdom:1985ag}
B.~Holdom, \emph{{Two U(1)'s and Epsilon Charge Shifts}},
  \href{https://doi.org/10.1016/0370-2693(86)91377-8}{\emph{Phys. Lett.}
  {\bfseries 166B} (1986) 196}.

\bibitem{Chang:2016ntp}
J.~H. Chang, R.~Essig and S.~D. McDermott, \emph{{Revisiting Supernova 1987A
  Constraints on Dark Photons}},
  \href{https://doi.org/10.1007/JHEP01(2017)107}{\emph{JHEP} {\bfseries 01}
  (2017) 107} [\href{https://arxiv.org/abs/1611.03864}{{\ttfamily
  1611.03864}}].

\bibitem{Hardy:2016kme}
E.~Hardy and R.~Lasenby, \emph{{Stellar cooling bounds on new light particles:
  plasma mixing effects}},
  \href{https://doi.org/10.1007/JHEP02(2017)033}{\emph{JHEP} {\bfseries 02}
  (2017) 033} [\href{https://arxiv.org/abs/1611.05852}{{\ttfamily
  1611.05852}}].

\bibitem{Rrapaj:2015wgs}
E.~Rrapaj and S.~Reddy, \emph{{Nucleon-nucleon bremsstrahlung of dark gauge
  bosons and revised supernova constraints}},
  \href{https://doi.org/10.1103/PhysRevC.94.045805}{\emph{Phys. Rev.}
  {\bfseries C94} (2016) 045805}
  [\href{https://arxiv.org/abs/1511.09136}{{\ttfamily 1511.09136}}].

\bibitem{An:2013yfc}
H.~An, M.~Pospelov and J.~Pradler, \emph{{New stellar constraints on dark
  photons}}, \href{https://doi.org/10.1016/j.physletb.2013.07.008}{\emph{Phys.
  Lett.} {\bfseries B725} (2013) 190}
  [\href{https://arxiv.org/abs/1302.3884}{{\ttfamily 1302.3884}}].

\bibitem{Braaten:1993jw}
E.~Braaten and D.~Segel, \emph{{Neutrino energy loss from the plasma process at
  all temperatures and densities}},
  \href{https://doi.org/10.1103/PhysRevD.48.1478}{\emph{Phys. Rev.} {\bfseries
  D48} (1993) 1478} [\href{https://arxiv.org/abs/hep-ph/9302213}{{\ttfamily
  hep-ph/9302213}}].

\bibitem{Siegert:2015knp}
T.~Siegert, R.~Diehl, G.~Khachatryan, M.~G.~H. Krause, F.~Guglielmetti,
  J.~Greiner et~al., \emph{{Gamma-ray spectroscopy of Positron Annihilation in
  the Milky Way}},
  \href{https://doi.org/10.1051/0004-6361/201527510}{\emph{Astron. Astrophys.}
  {\bfseries 586} (2016) A84}
  [\href{https://arxiv.org/abs/1512.00325}{{\ttfamily 1512.00325}}].

\bibitem{Prantzos:2010wi}
N.~Prantzos et~al., \emph{{The 511 keV emission from positron annihilation in
  the Galaxy}}, \href{https://doi.org/10.1103/RevModPhys.83.1001}{\emph{Rev.
  Mod. Phys.} {\bfseries 83} (2011) 1001}
  [\href{https://arxiv.org/abs/1009.4620}{{\ttfamily 1009.4620}}].

\bibitem{Beacom:2005qv}
J.~F. Beacom and H.~Yuksel, \emph{{Stringent constraint on galactic positron
  production}},
  \href{https://doi.org/10.1103/PhysRevLett.97.071102}{\emph{Phys. Rev. Lett.}
  {\bfseries 97} (2006) 071102}
  [\href{https://arxiv.org/abs/astro-ph/0512411}{{\ttfamily
  astro-ph/0512411}}].

\bibitem{Sizun:2006uh}
P.~Sizun, M.~Casse and S.~Schanne, \emph{{Continuum gamma-ray emission from
  light dark matter positrons and electrons}},
  \href{https://doi.org/10.1103/PhysRevD.74.063514}{\emph{Phys. Rev.}
  {\bfseries D74} (2006) 063514}
  [\href{https://arxiv.org/abs/astro-ph/0607374}{{\ttfamily
  astro-ph/0607374}}].

\bibitem{Groh:2013mma}
J.~H. Groh, G.~Meynet, C.~Georgy and S.~Ekstrom, \emph{{Fundamental properties
  of core-collapse Supernova and GRB progenitors: predicting the look of
  massive stars before death}},
  \href{https://doi.org/10.1051/0004-6361/201321906}{\emph{Astron. Astrophys.}
  {\bfseries 558} (2013) A131}
  [\href{https://arxiv.org/abs/1308.4681}{{\ttfamily 1308.4681}}].

\bibitem{2014MNRAS.437..532M}
M.~{Matsuura}, J.~A. {Yates}, M.~J. {Barlow}, B.~M. {Swinyard}, P.~{Royer},
  J.~{Cernicharo} et~al., \emph{{Herschel SPIRE and PACS observations of the
  red supergiant VY CMa: analysis of the molecular line spectra}},
  \href{https://arxiv.org/abs/1310.2947}{{\ttfamily 1310.2947}}.

\bibitem{Adams:2013ana}
S.~M. Adams, C.~S. Kochanek, J.~F. Beacom, M.~R. Vagins and K.~Z. Stanek,
  \emph{{Observing the Next Galactic Supernova}},
  \href{https://doi.org/10.1088/0004-637X/778/2/164}{\emph{Astrophys. J.}
  {\bfseries 778} (2013) 164}
  [\href{https://arxiv.org/abs/1306.0559}{{\ttfamily 1306.0559}}].

\bibitem{2010ApJ...725..940Y}
S.-C. {Yoon}, S.~E. {Woosley} and N.~{Langer}, \emph{{Type Ib/c Supernovae in
  Binary Systems. I. Evolution and Properties of the Progenitor Stars}},
  \href{https://doi.org/10.1088/0004-637X/725/1/940}{\emph{Astrophys. J.}
  {\bfseries 725} (2010) 940}
  [\href{https://arxiv.org/abs/1004.0843}{{\ttfamily 1004.0843}}].

\bibitem{Crowther:2006dd}
P.~A. Crowther, \emph{{Physical Properties of Wolf-Rayet Stars}},
  \href{https://doi.org/10.1146/annurev.astro.45.051806.110615}{\emph{Ann. Rev.
  Astron. Astrophys.} {\bfseries 45} (2007) 177}
  [\href{https://arxiv.org/abs/astro-ph/0610356}{{\ttfamily
  astro-ph/0610356}}].

\bibitem{Graur:2016lca}
O.~Graur, F.~B. Bianco, M.~Modjaz, I.~Shivvers, A.~V. Filippenko, W.~Li et~al.,
  \emph{{LOSS Revisited - II: The relative rates of different types of
  supernovae vary between low- and high-mass galaxies}},
  \href{https://doi.org/10.3847/1538-4357/aa5eb7}{\emph{Astrophys. J.}
  {\bfseries 837} (2017) 121}
  [\href{https://arxiv.org/abs/1609.02923}{{\ttfamily 1609.02923}}].

\bibitem{1993ApJ...405..614C}
K.-W. {Chan} and R.~E. {Lingenfelter}, \emph{{Positrons from supernovae}},
  \href{https://doi.org/10.1086/172393}{\emph{Astrophys. J.} {\bfseries 405}
  (1993) 614}.

\bibitem{FRASER196863}
P.~Fraser, \emph{Positrons and positronium in gases},  vol.~4 of \emph{Advances
  in Atomic and Molecular Physics}, pp.~63 -- 107, Academic Press, (1968),
  \href{https://doi.org/https://doi.org/10.1016/S0065-2199(08)60185-2}{DOI}.

\bibitem{refId0}
{Janka, H.-Th.}, {M\"uller, B.}, {Kitaura, F. S.} and {Buras, R.},
  \emph{Dynamics of shock propagation and nucleosynthesis conditions in o-ne-mg
  core supernovae},
  \href{https://doi.org/10.1051/0004-6361:20079334}{\emph{A\&A} {\bfseries 485}
  (2008) 199}.

\bibitem{Kazanas:2014mca}
D.~Kazanas, R.~N. Mohapatra, S.~Nussinov, V.~L. Teplitz and Y.~Zhang,
  \emph{{Supernova Bounds on the Dark Photon Using its Electromagnetic Decay}},
  \href{https://doi.org/10.1016/j.nuclphysb.2014.11.009}{\emph{Nucl. Phys.}
  {\bfseries B890} (2014) 17}
  [\href{https://arxiv.org/abs/1410.0221}{{\ttfamily 1410.0221}}].

\bibitem{Piran:1999kx}
T.~Piran, \emph{{Gamma-ray bursts and the fireball model}},
  \href{https://doi.org/10.1016/S0370-1573(98)00127-6}{\emph{Phys. Rept.}
  {\bfseries 314} (1999) 575}
  [\href{https://arxiv.org/abs/astro-ph/9810256}{{\ttfamily
  astro-ph/9810256}}].

\bibitem{Meszaros:2006rc}
P.~Meszaros, \emph{{Gamma-Ray Bursts}},
  \href{https://doi.org/10.1088/0034-4885/69/8/R01}{\emph{Rept. Prog. Phys.}
  {\bfseries 69} (2006) 2259}
  [\href{https://arxiv.org/abs/astro-ph/0605208}{{\ttfamily
  astro-ph/0605208}}].

\bibitem{PhysRevLett.62.509}
E.~W. Kolb and M.~S. Turner, \emph{Limits to the radiative decays of neutrinos
  and axions from $\ensuremath{\gamma}$-ray observations of sn 1987a},
  \href{https://doi.org/10.1103/PhysRevLett.62.509}{\emph{Phys. Rev. Lett.}
  {\bfseries 62} (1989) 509}.

\bibitem{1538-3873-104-679-717}
P.~Podsiadlowski, \emph{The progenitor of sn 1987a}, {\emph{Publications of the
  Astronomical Society of the Pacific} {\bfseries 104} (1992) 717}.

\bibitem{Beacom:2010kk}
J.~F. Beacom, \emph{{The Diffuse Supernova Neutrino Background}},
  \href{https://doi.org/10.1146/annurev.nucl.010909.083331}{\emph{Ann. Rev.
  Nucl. Part. Sci.} {\bfseries 60} (2010) 439}
  [\href{https://arxiv.org/abs/1004.3311}{{\ttfamily 1004.3311}}].

\bibitem{doi:10.1063/1.1303252}
K.~Watanabe, M.~D. Leising, G.~H. Share and R.~L. Kinzer, \emph{The mev cosmic
  gamma-ray background measured with smm},
  \href{https://doi.org/10.1063/1.1303252}{\emph{AIP Conference Proceedings}
  {\bfseries 510} (2000) 471}
  [\href{https://arxiv.org/abs/https://aip.scitation.org/doi/pdf/10.1063/1.1303252}{{\ttfamily
  https://aip.scitation.org/doi/pdf/10.1063/1.1303252}}].

\bibitem{0004-637X-475-1-361}
R.~L. Kinzer, G.~V. Jung, D.~E. Gruber, J.~L. Matteson,  and L.~E. Peterson,
  \emph{Diffuse cosmic gamma radiation measured by heao 1}, {\emph{Astrophys.
  J.} {\bfseries 475} (1997) 361}.

\bibitem{1996A&AS..120C.381S}
A.~W. {Strong}, K.~{Bennett}, H.~{Bloemen}, R.~{Diehl}, W.~{Hermsen},
  W.~{Purcell} et~al., \emph{{Diffuse galactic hard X-ray and low-energy
  gamma-ray continuum.}}, {\emph{Astron. Astrophys. Supp.} {\bfseries 120}
  (1996) 381}.

\bibitem{Fradette:2014sza}
A.~Fradette, M.~Pospelov, J.~Pradler and A.~Ritz, \emph{{Cosmological
  Constraints on Very Dark Photons}},
  \href{https://doi.org/10.1103/PhysRevD.90.035022}{\emph{Phys. Rev.}
  {\bfseries D90} (2014) 035022}
  [\href{https://arxiv.org/abs/1407.0993}{{\ttfamily 1407.0993}}].

\bibitem{Redondo:2008ec}
J.~Redondo and M.~Postma, \emph{{Massive hidden photons as lukewarm dark
  matter}}, \href{https://doi.org/10.1088/1475-7516/2009/02/005}{\emph{JCAP}
  {\bfseries 0902} (2009) 005}
  [\href{https://arxiv.org/abs/0811.0326}{{\ttfamily 0811.0326}}].

\bibitem{Essig:2013goa}
R.~Essig, E.~Kuflik, S.~D. McDermott, T.~Volansky and K.~M. Zurek,
  \emph{{Constraining Light Dark Matter with Diffuse X-Ray and Gamma-Ray
  Observations}}, \href{https://doi.org/10.1007/JHEP11(2013)193}{\emph{JHEP}
  {\bfseries 11} (2013) 193} [\href{https://arxiv.org/abs/1309.4091}{{\ttfamily
  1309.4091}}].

\bibitem{Jaeckel:2017tud}
J.~Jaeckel, P.~C. Malta and J.~Redondo, \emph{{Decay photons from the axionlike
  particles burst of type II supernovae}},
  \href{https://doi.org/10.1103/PhysRevD.98.055032}{\emph{Phys. Rev.}
  {\bfseries D98} (2018) 055032}
  [\href{https://arxiv.org/abs/1702.02964}{{\ttfamily 1702.02964}}].

\bibitem{Payez:2014xsa}
A.~Payez, C.~Evoli, T.~Fischer, M.~Giannotti, A.~Mirizzi and A.~Ringwald,
  \emph{{Revisiting the SN1987A gamma-ray limit on ultralight axion-like
  particles}}, \href{https://doi.org/10.1088/1475-7516/2015/02/006}{\emph{JCAP}
  {\bfseries 1502} (2015) 006}
  [\href{https://arxiv.org/abs/1410.3747}{{\ttfamily 1410.3747}}].

\bibitem{Chang:2018rso}
J.~H. Chang, R.~Essig and S.~D. McDermott, \emph{{Supernova 1987A Constraints
  on Sub-GeV Dark Sectors, Millicharged Particles, the QCD Axion, and an
  Axion-like Particle}},
  \href{https://doi.org/10.1007/JHEP09(2018)051}{\emph{JHEP} {\bfseries 09}
  (2018) 051} [\href{https://arxiv.org/abs/1803.00993}{{\ttfamily
  1803.00993}}].

\bibitem{doi:10.1063/1.1307028}
G.~Weidenspointner, M.~Varendorff, S.~C. Kappadath, K.~Bennett, H.~Bloemen,
  R.~Diehl et~al., \emph{The cosmic diffuse gamma-ray background measured with
  comptel}, \href{https://doi.org/10.1063/1.1307028}{\emph{AIP Conference
  Proceedings} {\bfseries 510} (2000) 467}
  [\href{https://arxiv.org/abs/https://aip.scitation.org/doi/pdf/10.1063/1.1307028}{{\ttfamily
  https://aip.scitation.org/doi/pdf/10.1063/1.1307028}}].

\end{thebibliography}\endgroup

\end{document}